# A lexicon obtained and validated by a data-driven approach for organic residues valorization in emerging and developing countries

Christiane Rakotomalala[a,*], Jean-Marie Paillat[b,c], Frédéric Feder[d], Angel Avadí[e], Laurent Thuriès[d], Marie-Liesse Vermeire[f], Jean-Michel Médoc[d], Tom Wassenaar[d], Caroline Hottelart[c], Lilou Kieffer[c], Elisa Ndjie[c], Mathieu Picart[c], Jorel Tchamgoue[c], Alvin Tulle[c], Laurine Valade[c], Annie Boyer[g], Marie-Christine Duchamp[g], Mathieu Roche[h]

[a] CIRAD, UPR Recyclage et Risque, F-97743 Saint-Denis, Réunion, France.

Recyclage et Risque, Univ Montpellier, CIRAD, Montpellier, France.

[b] CIRAD, UPR Recyclage et Risque, F-49000 Angers, France.

Recyclage et Risque, Univ Montpellier, CIRAD, Montpellier, France

[c] ISTOM, 4 Rue Joseph Lakanal, 49000 Anger France

[d] CIRAD, UPR Recyclage et Risque, F-34398 Montpellier, France.

Recyclage et Risque, Univ Montpellier, CIRAD, Montpellier, France

[e] CIRAD, UPR Recyclage et Risque, Abidjan, Côte d'Ivoire.

Recyclage et Risque, Univ Montpellier, CIRAD, Montpellier, France.

[f] CIRAD, UPR Recyclage et Risque, 18524 Dakar, Sénégal.

Recyclage et Risque, Univ Montpellier, CIRAD, Montpellier, France

[g] CIRAD, DGDRS, DIST, F-34398 Montpellier, France

[h] CIRAD, UMR TETIS, F-34398 Montpellier, France.

TETIS, Univ Montpellier, AgroParisTech, CIRAD, CNRS, INRAE, Montpellier, France.

**Abstract**
The text mining method presented in this paper was used for annotation of terms related to biological transformation and valorization of organic residues in agriculture in low and middle-income country. Specialized lexicon was obtained through different steps: corpus and extraction of terms, annotation of extracted terms, selection of pertinent terms.

**Background**

The global aim of the study is to establish the state-of-the-art in the valorization of organic residues in agriculture in developing countries. The method described herein represents the first stage of analyses which can be conducted on bibliographic corpus extracted from online databases. The text mining carried out in this work generates a specialized lexicon for Natural Language Processing (NPL) on a specific area. This exploratory methodology could be used to guide a more in-depth and oriented text analysis of scientific publications (i.e. scientometric analysis). Moreover, this methodology can be reused and/or adapted in other domain depending on purpose.

**Method details**

1. Construction of the corpus

Several online databases were consulted in 2021, to extract articles relating to biotransformation and valorization in agriculture of organic residues in emerging and developing countries (WoS, Ovid, Scopus, Google scholar, HAL, Cairn.info, AGRIS, and Agritrop[1]) published until 2021. Terms used for bibliographic search in WoS are grouped as follows:

TS = ("sewage sludge" OR "crop residue*" OR "agricultural waste" OR "industrial waste" OR "food waste" OR "household waste" OR "organic waste" OR "urban waste" OR "co-product*" OR "by-product*" OR "biomass" OR "organic waste product*" OR mulch OR digestate* OR compost*) AND TS = (decomposition OR fermentation OR anaerobic OR aerobic OR methanisation OR composting OR vermicomposting OR fertilization OR bokashi OR biodegradation OR mineralization OR recycling OR "agricultural valuation" OR biotransformation OR mulching) AND TS = (africa OR "acp countries" OR "central america" OR "south america" OR "latin america" OR "south east asia" OR "south asia" OR afghanistan OR angola OR albania OR argentina OR armenia OR antigua OR azerbaijan OR burundi OR benin OR "burkina faso" OR bangladesh OR bosnia OR belarus OR belize OR bolivia OR brazil OR bhutan OR botswana OR "central african republic" OR china OR "ivory coast" OR cameroon OR congo OR colombia OR comoros OR "cape verde" OR "costa rica" OR cuba OR djibouti OR dominica OR "dominican republic" OR algeria OR ecuador OR egypt OR eritrea OR ethiopia OR fiji OR micronesia OR gabon OR georgia OR ghana OR guinea OR gambia OR grenada OR guatemala OR guyana OR honduras OR haiti OR indonesia OR india OR iran OR iraq OR jamaica OR jordan OR kazakhstan OR kenya OR kyrgyzstan OR cambodia OR kiribati OR "lao people's democratic republic" OR lebanon OR liberia OR libya OR "saint lucia" OR "sri lanka" OR lesotho OR morocco OR moldova OR madagascar OR maldives OR mexico OR "marshall islands" OR "north macedonia" OR mali OR myanmar OR montenegro OR mongolia OR mozambique OR mauritania OR montserrat OR mauritius OR malawi OR malaysia OR namibia OR niger OR nigeria OR nicaragua OR niue OR nepal OR nauru OR pakistan OR panama OR peru OR philippines OR palau OR "papua new guinea" OR " Democratic People's Republic of Korea" OR "north korea" OR paraguay OR "palestinian territory" OR rwanda OR sudan OR senegal OR "saint helena, ascension and tristan da cunha" OR "solomon islands" OR "sierra leone" OR "el Salvador" OR somalia OR serbia OR "south sudan" OR "sao tome and principe" OR suriname OR eswatini OR "syrian arab republic" OR chad OR togo OR thailand OR tajikistan OR tokelau OR turkmenistan OR "timor-leste" OR tonga OR tunisia OR turkey OR tuvalu OR tanzania OR uganda OR ukraine OR uzbekistan OR "saint vincent and the grenadines" OR venezuela OR "vietnam" OR vanuatu OR "wallis and futuna" OR samoa OR yemen OR "south africa" OR zambia OR zimbabwe).

This equation was thereafter adapted for the other databases specificities. The search gave 24 186 references on which a selective sorting was conducted to avoid duplicates and to select references in English only. A total of 7 692 references were used to generate the dataset available in the excel file (Initial_Corpus_References.xlsx ) available on depository [1]. The corpus of the dataset combines articles, reports, book sections, and student thesis with bibliographic references (authors, year of publication, title, doi, and url).

2. Extraction of candidate terms

BioTex [5] was used to perform an Automatic Term Extraction (ATE) on the corpus. The terms extracted (e.g. rumen, humic acid, nutrient recovery, …) give a semantic point of view of the theme of the text. This tool was developed for Biomedical term extraction [2] and was adapted to extract terms associated with food security [3]. First, BioTex performed a linguistic screening through syntactic patterns (noun-noun, adjective-noun, …). In order to rank terms extracted on the "titles" corpus, the F-TF-IDF-C score integrated to BIOTEX was applied. This measure combines (i) C-value [4] to favor multi-word terms extracted, and (ii) TF-IDF (Term Frequency-Inverse Document Frequency) to highlight discriminative terms [2].

---

[1] http://webofknowledge.com/WOS - https://ovidsp.ovid.com/autologin - https://scholar.google.com/ - https://hal.science/ - https://www.cairn.info/ - https://agris.fao.org/ - http://agritrop.cirad.fr/

Text mining was thereafter performed on titles of the corpus using the BioTex tools [5] and the result can be found in the associated excel file (Extracted_Terms.xlsx) on depository [1]. The first column contains the 19 580 terms obtained from the extraction. The second column ("term") presents the terms constituted of words or compound nouns (e.g. mulch, effluents, soil amendments, bagasse co-composting). The rank, in the last column, is obtained by maximizing a discriminative score associated with terms (i.e. F-TF-IDF-C).

**Method validation**

1. Annotation guide V1 and associated Fleiss Kappa

Five specialist raters conducted a first annotation on 200 sampled candidate terms among the 19 580 to exclude irrelevant terms to the topic of interest following the guideline file (Annotation_guidelines.pdf) available on the depository [1]. Specialists are researchers working on organic residues valorization in agriculture. Each rater was asked to categorize each candidate term belonging to i) organic residues (OWT) and/or ii) biotransformation process (TM) and/or iii) valorization in agriculture (AV) or iv) none of them (None) following the first annotation guide. Definition of each category is described in the annotation guide. Table 1 shows example of the first step of annotation conducted by specialist.

*Table 1: Examples of annotation process*

| Term | Expert 1 | | | Expert 2 | | | Expert 3 | Expert 4 | | Expert 5 | |
| --- | --- | --- | --- | --- | --- | --- | --- | --- | --- | --- | --- |
| | Category 1 | Category 2 | Category 3 | Category 1 | Category 2 | Category 3 | Category 1 | Category 1 | Category 2 | Category 1 | Category 2 |
| manure | OWT | TM | AV | OWT | | | OWT | OWT | AV | OWT | |
| anaerobic digestion | TM | | | TM | | | TM | TM | | TM | |
| biogas | TM | | | TM | | | None | None | | TM | |
| rice | AV | | | OWT | AV | | OWT | None | | AV | OWT |
| nitrogen | OWT | TM | AV | OWT | TM | AV | None | None | | AV | TM |

The Fleiss Kappa [6] which measures agreement between several raters equals to 0.52 for this first annotation corresponding to a bad agreement between the 5 raters. The 4 categories chosen to annotate the candidate terms appeared to be too restrictive. Terms indirectly associated to one or more of the 4 categories have been excluded by several raters.

2. Annotation guide V2 and associated Fleiss Kappa

In a second annotation guideline, the manual labelling process focuses on the overall degree of pertinence related to the topic of valorization of organic residues. In this context, candidate term was annotated according 3 classes: (i) very pertinent when it was directly connected to one or more category(ries) (i.e. OWT+, TM+, AV+), (ii) pertinent when it was indirectly connected to one or more category(ries) (i.e. OWT, TM, AV), and (iii) irrelevant (i.e. None).

A second annotation on the same 200 sampled terms was conducted. Fleiss Kappa was calculated for 3 and 5 raters. It revealed a decreasing trend of the value (0.84 to 0.60) with increasing number of raters. Closer comparison highlighted more terms indirectly related to one or more category(ies) selected by 3 raters with high value of Kappa. In order to include as many terms indirectly related to the subject as possible, it was decided to apply the logic of these 3 annotators to pursue the categorization of the remaining terms.

In Table 2, the results are evaluated in terms of precision (percentage of pertinent terms) obtained over the top k extracted terms (P@k). The results confirm that the ranking function of BioTex is relevant by highlighting relevant terms at the top of the list. The above detailed dataset can be found in the CIRAD Dataverse repository [1].

*Table 2: Precision (P@k) according the BioTex ranking*

| Rank of term - k | Precision – P@k (%) |
|---|---|
| 100 | 83 |
| 200 | 81 |
| 500 | 59 |
| 1000 | 49 |
| 3000 | 41 |
| 5000 | 36 |
| 10000 | 20 |
| 19580 | 25 |

3. Annotation and validation of all extracted terms

One of the 5 specialists then pursued the annotation, with a degree of relevance, on the remaining extracted terms. It was decided to continue the categorization with the degree of pertinence and to apply the logic of the 3 annotators with the high kappa value explained above. It took about one-week work for the rater to conduct the categorization. The same five raters were then asked to verify and finalize the terms selection related to the biotransformation and valorization in agriculture of organic residues in low-income countries. All verified relevant terms were combined in the last file on the depository (Pertinent_Terms.xlsx), containing terms which can be indirectly (first sheet) or directly (second to fourth sheet) related to the topic.

From the 19 580 initial candidate terms, about 75% were not associated to the topic of interest (Table 2). Irrelevant terms included words which are not related to organic residues nor biotransformation nor valorization in agriculture, such as: absence, certification, design, effect, fecundity, fitness, gray, immune response, integration analysis, low cost, marker genes, …. Among the 25% relevant terms, 2 079 were closely associated with the organic residues valorization in emerging and developing countries such as sludge, sewage, livestock, manure, slurry, anaerobic digestion, composting, vermicomposting. Several terms can be found in the glossary of terms related to livestock and manure management [7] and figure among terms with high pertinence in this dataset. Moreover, some of relevant terms are cited in literatures as the biotransformation (e.g.: anaerobic digestion, composting, bioethanol, biohydrogen) and valorization in agriculture (e.g.: biofertilization, organic fertilizers, amendments) of organic residues (e.g. rice straw, sugarcane bagasse, animal manure) [8], [9].

**Discussions**

The text-mining tool used in this work is based on statistical criteria that highlight discriminative terms. This method identifies significant terms that are present in the texts. As future work, the proposed framework could be extended by extracting variation of terms [10] that enables to recognize rare and/or unsystematic terms but also synonyms. Moreover embedding approaches [11] and generative methods based on LLM (Large Language Models) techniques [12] could be applied to recognize new terms.

**Acknowledgments**

This work was supported by European Union Feder funding (EU ERDF: GURTDI 20151501-0000735).